\documentclass[twocolumn,showpacs,preprintnumbers,amsmath,amssymb, superscriptaddress,prb,aps]{revtex4-1}
\usepackage{eqnarray}
\usepackage{amsmath}
\usepackage{amssymb}
\usepackage{graphicx}
\usepackage{dcolumn}
\usepackage{bm}

\usepackage{color}
\definecolor{red}{rgb}{1,0,0}

\definecolor{blue}{rgb}{0,0,1}

\definecolor{green}{rgb}{0,1,0}

\begin{document}
\preprint{APS}
\author {Mariano de Souza}
\email{mariano@rc.unesp.br; Present Address: Institute of Semiconductor and Solid State Physics, Johannes Kepler University Linz, Austria.}
\author {Ricardo Paupitz}
\author {Antonio Seridonio}
\author {Roberto E. Lagos}
\affiliation{IGCE, Unesp - Univ Estadual Paulista, Departamento de F\'{i}sica, 13506-900, Rio Claro, SP, Brazil}
\title{Specific Heat Anomalies in Solids Described by a Multilevel Model}
\vspace{0.5cm}
\begin{abstract}


Specific heat measurements constitute one of the most powerful
experimental methods to   probe fundamental excitations in
solids. After the proposition of
Einstein's model, more than one century ago  (Annalen der Physik \textbf{22}, 180 (1907)), several theoretical models have
been proposed to describe experimental results. Here we report on a
detailed analysis of the two-peak specific heat anomalies observed in
several materials. Employing a simple multilevel model, varying
the spacing between the energy levels  $\Delta_i$ = $(E_i$ $-$
$E_{0})$ and the degeneracy of each energy level $g_i$, we derive
the required conditions for the appearance of such anomalies.  Our
findings indicate that a ratio of $\Delta_2$/$\Delta_1$
$\thickapprox$ 10 
between the energy levels and a high degeneracy of one of the energy levels
define the two-peaks regime in the specific heat. 
Our approach accurately matches recent
experimental results. Furthermore, using a mean-field approach we calculate the specific heat of a degenerate Schottky-like system undergoing a ferromagnetic (FM) phase transition. Our results reveal that as the degeneracy is increased the Schottky maximum in the specific heat becomes narrow while the peak associated with the FM transition remains unaffected.
\end{abstract}


\maketitle


\date{\today}
\section{Introduction}

In the field of condensed matter Physics, specific heat measurements can be considered
as a pivotal experimental technique for characterizing the
fundamental excitations involved in a certain phase transition. Indeed, phase
transitions involving spin \cite{desouza2009,PhysRevLett.104.016403}, charge \cite{PhysRevB.82.144438}, lattice \cite{PhysRevB.81.134525} (phonons) and orbital degrees of freedom, the interplay between ferromagnetism and superconductivity \cite{PhysRevB.86.020501}, Schottky-like anomalies in doped compounds \cite{alzira}, electronic levels in finite correlated systems \cite{lazaro}, among other features, can be captured by means of high-resolution
calorimetry. Furthermore, the entropy change associated with a
first-order phase transition, no matter its nature, can be directly
obtained upon integrating the specific heat over $T$,
i.e.\,$C$($T$)/$T$, in the temperature range of interest. In his seminal paper of 1907 \cite{Einstein1907},
in order to explain the deviation of the
specific heat of certain materials, like silicon, boron and carbon,
from the Dulong-Petit's law, Einstein proposed a model based on the
assumption that all atoms in a solid vibrate independently from each
other with the same eigen-frequency. The well-known Einstein's expression for the phononic specific heat at constant volume, $C^{E}_{ph,v}(T)$, is  given  by:
\begin{equation}
C^{E}_{ph,v}(T) = (\beta h \nu)^2  \frac{e^{\beta h \nu}}{(e^{\beta h \nu} - 1)^2},
 \label{Einstein}
\end{equation}
hereafter all extensive quantities are defined as per particle, except when otherwise indicated;
$\beta$ = 1/$T$,  ($k_B$ = 1 hereafter), $\nu$ is the eigen-frequency of the oscillator and $h$ the Planck's constant. The quantity $h \nu$ is the so-called Einstein temperature $\Theta_E$.
It is worth mentioning that recently, one of us made use of Einstein's model to determine the eigen-energy of the counter-anions libration modes in a molecular conductor \cite{Foury2013}.
\begin{figure}
\begin{center}
\includegraphics[angle=0,width=0.50\textwidth]{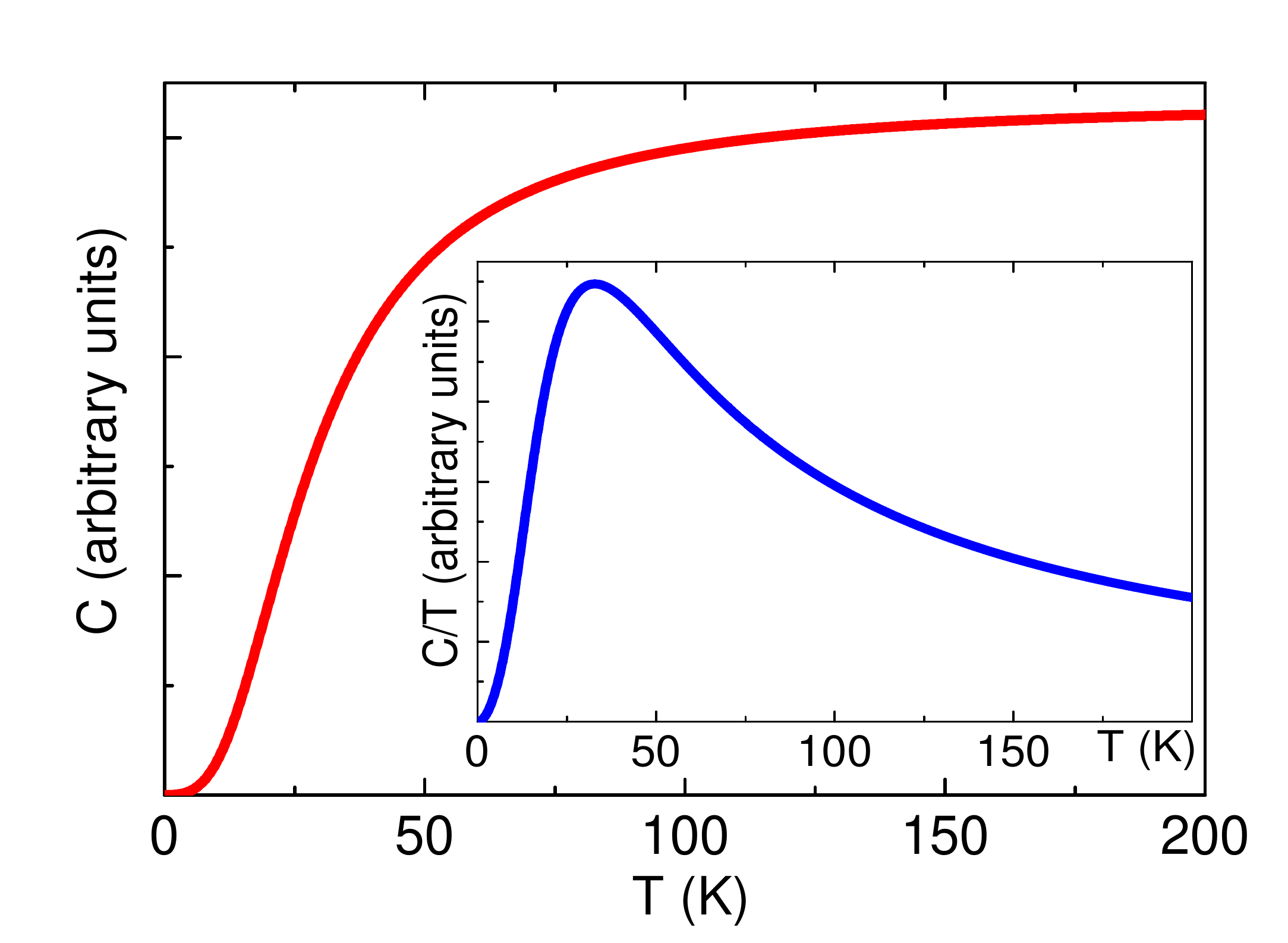}
\end{center}
\caption{\small Main panel: Specific heat ($C$) as a function of
temperature for an hypothetical \emph{Debye's solid} (cf.\,Eq.\,\ref{Debye}) with $\Theta_D$ = 120\,K.
Inset: $C$/$T$ versus $T$ showing a maximum centered at $T \simeq$ 0.28$\cdot
\Theta_D$.}\label{Debfig}
\end{figure}
However, thought the success of the model proposed by Einstein, the theoretical description introduced by P.\,Debye in 1912\cite{Debye1912}, constitutes the hallmark in the description of the phononic contribution to the specific heat in solids, see e.g.\,\cite{Gopal}.
Essentially, Debye's model is based on the hypothesis of a continuous
isotropic solid. The dispersion relation is linear,
i.e.\,the sound velocity is constant and isotropic, being a set of eigen-frequencies allowed to the oscillators. In the frame of
Debye's model, the phononic specific heat at constant volume,
$C^{D}_{ph,v}(T)$, reads:
\begin{equation}
C^{D}_{ph,v}(T) = 3\left( \frac{T}{\Theta_D}\right)^3
\int_0^{x}{\frac{x^4 e^x}{(e^x-1)^2}}dx, \label{Debye}
\end{equation}
where  $x = \Theta_D/T$ and $\Theta_D$ is the so-called
Debye temperature.  The behavior of the specific heat as a
function of $T$, according to Eq.\,\ref{Debye}, well-known from
textbooks, is shown in the main panel of Fig.\,\ref{Debfig}.
A particular behavior, usually not explicitly shown and discussed in textbooks \cite{pathria}, is observed by plotting $C/T$ versus $T$,
cf.\,inset of Fig.\,\ref{Debfig}. The maximum centered at 0.28$\cdot
\Theta_D$ corresponds to the inflexion point of Eq.\,\ref{Debye} and manifests itself as a direct effect of the accessible energy levels upon increasing the temperature. In other words, as the temperature of a certain \emph{Debye's solid} is increased the number of accessible energy levels is reduced and, as a consequence, the entropy variation rate is reduced.
The latter indicates that for \emph{any} system which obeys a
Debye-like behavior, $\Theta_D$ can be directly estimated by
plotting $C/T$ versus $T$. Similarly in Einstein's model (Eq.\,\ref{Einstein}), $C/T$ versus $T$
has a  maximum centered at 0.38$\cdot
\Theta_E$. A combined system of Debye and Einstein phonons can be studied provided the Einstein and Debye temperature scales are not so close  \cite{ajp2005}.

At low temperatures, namely for ordinary metals $T$ $\simeq$ $\Theta_D$/50  $\simeq$ liquid
$^4$He temperature, the following relation is valid \cite{Gopal}:
\begin{equation}
C_v(T) =  \gamma T + A T^3, \label{e-p}
\end{equation}
where $\gamma$ is the Sommerfeld coefficient and $A$ = 12$\pi^4
$/5$\Theta_D^3$. Here the Debye temperature can be directly
obtained via $A$ parameter, whereas the
effective mass $m^*$ of the charge carriers in a metal can be
estimated by means of the Sommerfeld coefficient \cite{Gopal}. The latter can be considered the
\emph{smoking gun}, for instance,  when discovering materials with
heavy-fermion-like behavior. Moreover, still considering the relevance of high-resolution calorimetry, combining $C_p$($T$) and the linear thermal expansion coefficient \cite{Review}, $\alpha_i$($T$), and making use of the
Ehrenfest theorem, see e.g.\,\cite{Gopal}, the uniaxial-pressure dependence of the critical
temperature for pressure applied along the $i$-axis for a
second-order phase transition can be directly obtained, as follows:
\begin{equation}
\left(\frac{dT_c}{dP_i}\right)_{P_i \rightarrow 0} =  V_{mol}
\cdot T_c \cdot \frac{\Delta\alpha_i}{\Delta C},\label{Ehrenfest}
\end{equation}
where $V_{mol}$ is the molar volume, $\Delta\alpha_i$ and $\Delta C$
refer to the thermal expansion and specific heat  (per mol)  jumps at the
transition temperature, respectively. The index \emph{i} refers to
the crystallographic direction, along which pressure is applied.
Rigorously speaking, the Ehrenfest relation is applicable only for
mean-field-like phase transition, where both $\Delta\alpha$ and $\Delta
C$ present step-like behavior. Note that the Ehrenfest relation
enables us to determine the pressure dependence of $T_c$ purely via
thermodynamic quantities, i.e., $dT_c$/$dP_i$ can be estimated
without carrying out any experiment under application of external
pressure.

For a general two-level system, separated by an energy gap $\Delta_1$, the specific heat is described by the following expression:
\begin{equation}\label{Schottky}
C = \frac{(\beta \Delta_1)^2e^{-\beta \Delta_1}}{(1 + e^{-\beta \Delta_1})^2},
\end{equation}
In such a system, the so-called Schottky anomaly manifests itself as a shallow maximum in the specific heat data as a function of temperature.


After this brief introduction, recalling some fundamental aspects related to specific heat measurements in solids, well-known from text books, see e.g.\,\cite{Gopal}, below we present a multilevel model to describe systems which present multiple peaks in the specific heat data. Interestingly enough, such a simple model is capable to describe experimental electronic specific heat results of correlated electrons systems.

\begin{figure}
\centering
\includegraphics[angle=0,width=0.28\textwidth]{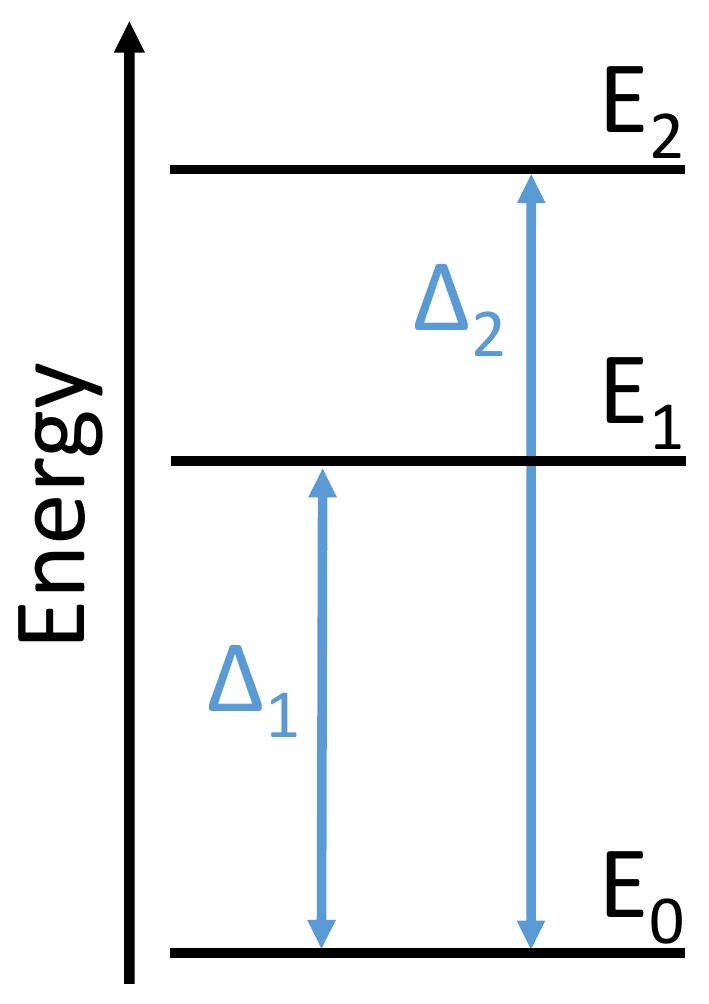}
\caption{Schematic representation of the energy levels $E_0$, $E_1$ and $E_2$ 
considered in the model (Eq.\,\ref{Z}). The excitation gaps $\Delta_1$ and $\Delta_2$ 
are defined as the energy difference
relative to the ground state energy ($E_0$). See discussion in the main text for
details.} \label{energy_levels}
\end{figure}

\begin{figure*}
\begin{center}
\includegraphics[scale=0.33]{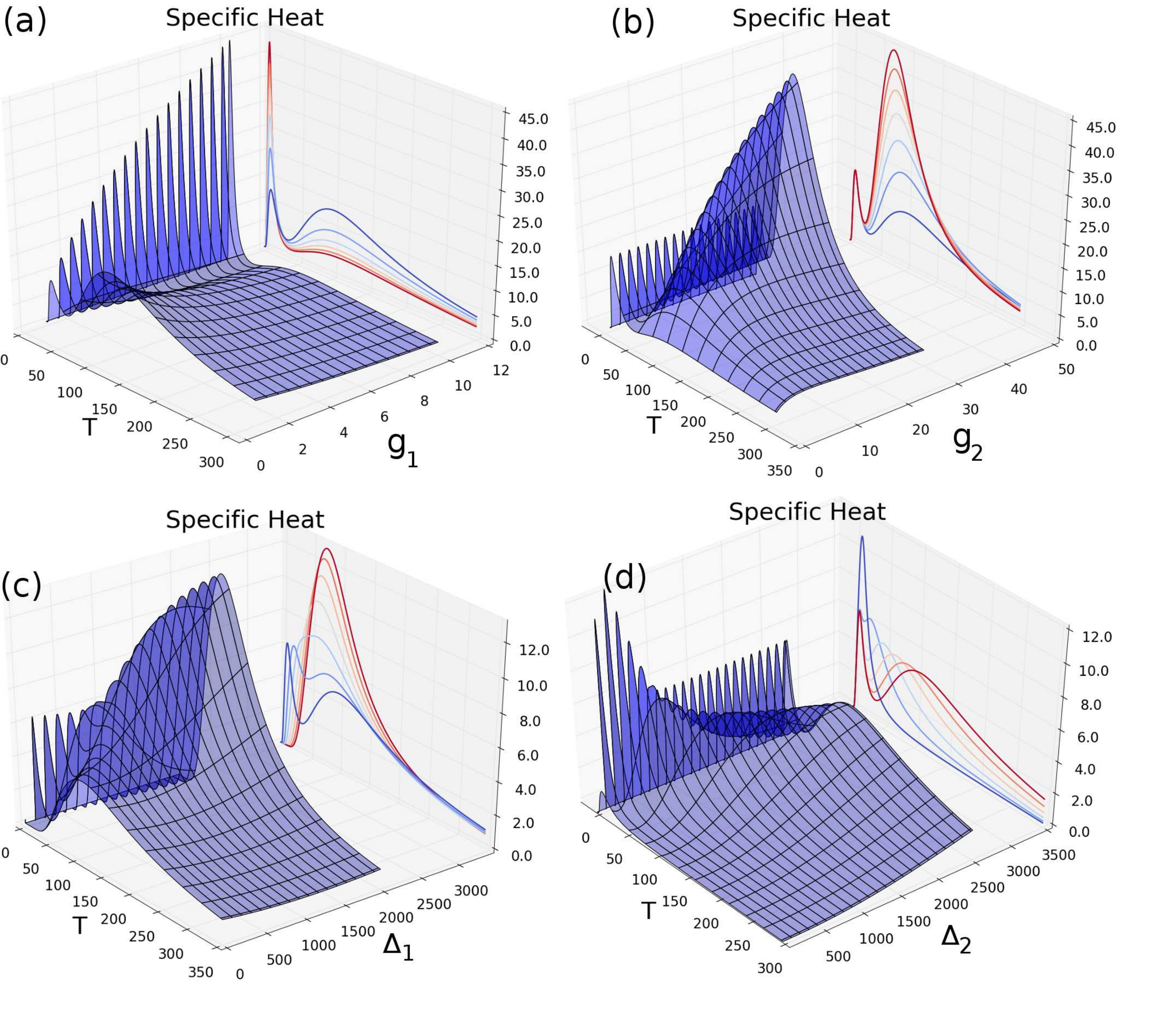}
\end{center}
\caption{\small Specific heat ($C$) as a function of:
(a) Temperature ($T$) and degeneracy of the energy level 1, labelled $g_1$; (b) $T$  and degeneracy of the energy level 2, $g_2$;
(c) $T$ and energy gap $\Delta_1$; (d) $T$ and energy gap $\Delta_2$. The projection of each curve shown in the 3D plots is depicted in the various panels. Details are discussed in the main text.}\label{fig_3d}
\end{figure*}


\section{Mathematical Modeling}
In the frame of the \emph{canonical ensemble} the sum over states, namely the partition function ($Z$) of a system, is given by:

\begin{equation}
\label{Z-original}
Z \equiv Tr (e^{-\beta \hat{H}})  = \sum_n e^{-\beta E_n},
\end{equation}
where $E_n$ refers to the $n^{th}$ energy level, i.e.\,the $n^{th}$ eigenvalue of the system's Hamiltonian $\hat{H}$ \cite{landau2013statistical}.
As a matter of fact,  the partition function encodes all information of the physical system, so that once we know the energy eigenvalues of the physical system all thermodynamic observables can be calculated. This in turn is not the case if one has the system Hamiltonian in hand \cite{Laughlin04012000}, being necessary to make its diagonalization to obtain $Z$ and then the physical quantities of interest.
For the sake of simplicity, in order to describe the two-peak like anomalies observed experimentally in several materials (see below), we propose a multilevel model approach.
This \emph{Ansatz} enables us to obtain directly the thermodynamic quantities of interest. In other words, within such an approach we work in the diagonal representation of the Hamiltonian $\hat{H}$ (cf.\,Eq.\,\ref{Z-original}).
We start by considering a three-level model\cite{meijer1973note}, with $\Delta_1 = (E_1-E_0)$ and $\Delta_2=(E_2-E_0)$, 
being $E_0$, $E_1$ and $E_2$ 
defined as shown in Fig.\,\ref{energy_levels}, the partition sum reads

\begin{equation}
Z = g_0e^{-\beta E_0}+g_1e^{-\beta \left(E_0+\Delta_1\right)}+g_2e^{-\beta \left(E_0+\Delta_2\right)}\label{Z},
\end{equation}
where $g_i$ indicates the degeneracy of each energy level ($i = 0,1,2$). The assumption that the energy gaps $\Delta_i$, i.e.\,energy scales for a generic system, do not depend on the temperature is quite realistic. In this regard we refer, for instance, to the Schottky model \cite{Gopal}, where the energy gap $\Delta_1$ separating the two energy-levels is fixed and thus temperature independent.

Making use of Eq.\,\ref{Z}, one can calculate the specific
heat, employing the following well-known relations

\begin{equation}
 E=-\frac{\partial}{\partial\beta}\left(ln{Z}\right),
 \label{energia}
\end{equation}

and

\begin{equation}
 C=-\beta^2\left(\frac{\partial E}{\partial\beta}\right),
\end{equation}

obtaining

\begin{equation}
  \label{specific-heat-function}
    C=\beta^2\frac{\sum_{i}\sum_{j} g_ig_j{\Delta_i}{\left(\Delta_i-\Delta_j\right)}e^{-\beta(\Delta_i+
      \Delta_j)}}{\left[\sum_{i} g_ie^{-\beta\Delta_i}\right]^2},
\end{equation}
where $i, j = 0, 1, 2$ and $\Delta_0 = 0$. Note that considering $j$ = 1 in Eq.\,\ref{specific-heat-function}, the specific heat for the Schottky model (Eq.\,\ref{Schottky}) is nicely restored.

For the sake of completeness, we calculated the summations in Eq.\,\ref{specific-heat-function} and present below the expression employed in our analysis:


\begin{align}
C & =\beta^{2}\dfrac{g_{1}g_{0}\Delta_{1}^{2}e^{-\beta\Delta_{1}}+g_{0}g_{2}\Delta_{2}^{2}e^{-\beta\Delta_{2}}}{[g_{0}+g_{1}e^{-\beta\Delta_{1}}+g_{2}e^{-\beta\Delta_{2}}]^{2}}+\nonumber \\
 & +\frac{g_{1}g_{2}e^{-\beta(\Delta_{1}+\Delta_{2})}[\Delta_{1}(\Delta_{1}-\Delta_{2})+\Delta_{2}(\Delta_{2}-\Delta_{1})]
}{[g_{0}+g_{1}e^{-\beta\Delta_{1}}+g_{2}e^{-\beta\Delta_{2}}]^{2}}.
\label{specific-heat-function1}
\end{align}


In the present model, the crucial feature refers to
the relation between the energy scales of the two excitation gaps. The latter together with the degeneracies $g_i$ of the energy levels,
define whether the system presents a Schottky-like behavior with a
single peak/maximum in the specific heat or if it will exhibit two or even three (in this case a four energy levels is required) peaks or maxima. More specifically, one can say that in such cases two and three energy scales govern the Physics of the system of interest.

\section{Results and Discussion}
In what follows, employing Eq.\,\ref{specific-heat-function1},  we discuss the required conditions for
the emergence of the two-peak anomalies in the specific heat. In
Fig.\,\ref{fig_3d} we present  $3$-dimensional (3D) plots of the
specific heat as a function of temperature and degeneracy of the energy levels
1 (Fig.\,\ref{fig_3d}-a) 
and 2 (Fig.\,\ref{fig_3d}-b) and, specific heat as a function of
temperature and the relative value of the energy levels $1$ 
(Fig.\,\ref{fig_3d}-c) and $2$
(Fig.\,\ref{fig_3d}-d) as well. The optimized parameters used in such set of fits, appropriate to give rise to the emergence of a double peak in the specific heat are $\Delta_0$ = 0, $\Delta_1$ = 205\,J/mol, $\Delta_2$ = 2327\,J/mol, $g_0$ = 1, $g_1$ = 2, $g_2$ = 4, i.e.\,in each 3D plot one of these parameters was varied while the remaining ones were kept constant. For instance, in Fig.\,\ref{fig_3d}-a) $T$ and $g_1$ were used as the variable parameters. Note that upon increasing the degeneracy $g_1$ the double peak in the specific heat gradually vanishes. A distinct situation is depicted in Fig.\,\ref{fig_3d}-b), where the double peak in the specific heat appears as the degeneracy $g_2$ is increased. In Fig.\,\ref{fig_3d}-c) and d) we vary the temperature and the energy levels $\Delta_1$ and $\Delta_2$, respectively. The present findings indicate two distinct ways of catching a double-peak in the specific heat, i.e.\,either by varying the degeneracy or by increasing the energy levels $\Delta_1$ or $\Delta_2$. More specifically, Figs.\,\ref{fig_3d}-a)
and b) indicate that at least one of the energy levels should be highly degenerated to give way to the double-peak in the specific heat. Similarly to the Schottky-like anomaly, this behavior suggest that the high degeneracy contribute dramatically to the specific heat over a quite restricted $T$-window.  Such a behavior can be easily understood in terms of the system entropy: once the number of accessible states is increased and this is the physical situation of a highly degenerated generic system or energy level, the entropy of the system is increased and, as a consequence, a double-peak structure shows up in the specific heat, since the entropy of the system is proportional to the area and can be estimated via $\Delta S = \int_{T_a}^{T_b} C/T dT$, where $T_a$ and $T_b$ indicate two distinct temperatures in the range of interest. In other words, the system entropy is increased at the expense of the satellite peak in the specific heat.
Nevertheless, as discussed above, the specific heat is a bulk property. Thus in order to map the energy levels of a generic system and its degeneracy, microscopic experimental techniques like electron spin and magnetic resonance \cite{abragam1970electron}, for instance, are required.
We stress that the two-maxima observed in the specific heat, obtained from the above-discussed \emph{Ansatz},
is a universal signature of \emph{any} system with two distinct characteristic energy scales
that differ from each other at least by roughly one order of magnitude. In fact,
such features in the specific heat are related to various ongoing topics of interest
in condensed matter physics. In particular, 4$f$-electrons based magnetic systems  \cite{RevModPhys.81.1551} show spectacular
properties in this regard.

\begin{figure}
\centering
\includegraphics[angle=0,width=0.50\textwidth]{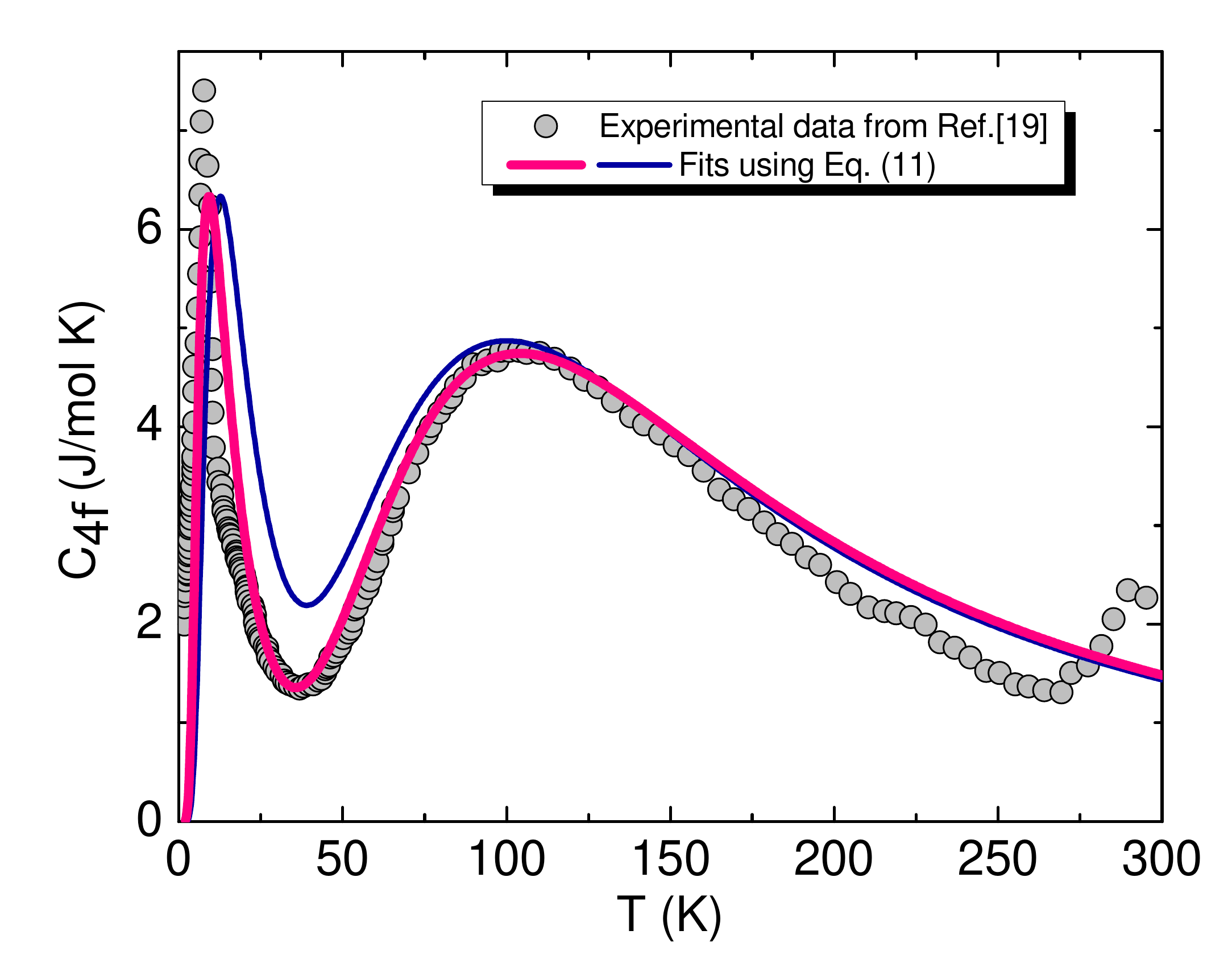}
\caption{\small Electronic specific heat experimental data $C_{4f}$ (circles) and theoretical fits (solid lines) for CeAuGe \cite{SondeziMhlungu20093032} using Eq.\,\ref{specific-heat-function1} with $k_B$ = 8.314\,J/(mol.K), $g_0$ = 1, $g_1$ = 2, $g_2$ = 4, $\Delta_1$ = 205\,J/mol (282\,J/mol) for the solid pink line (solid blue line)  and $\Delta_2$ = 2327\,J/mol. 
}
\label{Fits-4levels}
\end{figure}

Hereafter we focus on examples of
various materials classes, in which a two-peak like anomaly in the
specific heat versus temperature have been observed experimentally.
We start with the heavy-fermion compound Ce$_3$Pd$_{20}$Si$_6$, a magnetic
field-induced quantum critical system \cite{Custers2012189}. For
this material, J.\,Custers \emph{et al.}\, observed the presence of
two distinct peaks centered at $T_Q \simeq$ 0.5\,K and $T_N \simeq$
0.25\,K under an external magnetic field of 0.5\,T in the electronic
specific heat. These authors attributed the features at $T_Q$ and
$T_N$ to a antiferro-quadrupolar magnetic order of the Ce 4$f$
orbitals located on the  8$c$ site and to an antiferromagnetic order,
respectively. Yet, the authors point out that this scenario is
compatible with the $\Gamma_8$ quartet and $\Gamma_7$ doublet ground
states due to the difference of the crystal-field splitting of the
Ce atoms on the 4$a$ and 8$c$ lattice sites \cite{Goto2009024716}.  The
small energy difference of only $\Delta T \simeq$ 0.25\,K $\simeq$
20\,$\mu$eV between the transition temperatures $T_Q$ and $T_N$
suggests that for Ce$_3$Pd$_{20}$Si$_6$ a subtle change in the
system total energy is enough to change the character of the Ce 4$f$
orbitals.  Interestingly enough, the suppression of $T_Q$ gives rise
to a magnetic field induced quantum phase transition.
Also for CeAuGe, CeCuGe, CeCuSi
\cite{SondeziMhlungu20093032}, TbPO$_4$ \cite{Hill197817},
$A_3$Cu$_3$(PO$_4$)$_4$ ($A$ = Ca, Sr) and
Cu(3--Chloropyridine)$_2$(N$_3$)$_2$ \cite{PhysRevB.65.214418},
Cu(en)$_2$Ni(CN)$_4$ \cite{CoordinationChemistryReviews22451},
URu$_2$Si$_2$  \cite{PhysRevLett552727} where a magnetic transition
followed by the appearance of superconductivity is observed,
(Ce$_{1-x}$La$_x$)$_3$Al \cite{PhysRevB.55.5937}, PrOs$_4$Sb$_{12}$
\cite{PhysRevLett.90.057001} and UPt$_3$ \cite{PhysRevLett.62.1411}
with two superconducting transitions, double peaks in the specific
have been reported. We now present the results of fits that can be
compared to experimental data set for the specific heat
of a particular heavy fermion compound.
In Fig.\ref{Fits-4levels} we show experimental literature results of the electronic specific heat for CeAuGe \cite{SondeziMhlungu20093032} together with theoretical fits employing Eq.\,\ref{specific-heat-function1}. It is worth mentioning that the ordinary phononic contribution to the specific heat were subtracted by the authors of Ref.\,\cite{SondeziMhlungu20093032}, being thus shown in Fig.\ref{Fits-4levels} the bare electronic specific heat contribution originated from the 4$f$-electrons ($C_{4f}$). Interestingly enough,
despite the \emph{non-interacting gas} picture we have adopted in this work, the present approach is capable to describe an electronic system, in which correlation effects are present as discussed above.
In order to determine the nature of the electronic/magnetic ``excitations'' responsible for the emergence of a double-peak anomaly in the specific heat data of CeAuGe, as well for the various materials above-mentioned, microscopic/spectroscopic data are required and constitute a topic out of the scope of the present work. It is worth mentioning that
inelastic neutron scattering investigations \cite{SondeziMhlungu20093032}, carried out on CeAuGe at 15\,K, revealed the presence of pronounced crystal field excitations at 24.3\,meV ($\simeq$ 282\,K), which roughly corresponds to one of the energy scales, namely $\Delta_1$ = 205\,J/mol, employed in one of the fits (solid pink line) depicted in Fig.\,\ref{Fits-4levels}. Note that a fit using $\Delta_1$ = 282\,J/mol (solid blue line in Fig.\,\ref{Fits-4levels}) does not deviate so much from the experimental data.  Hence, we figure out that the energy scales $\Delta_i$, employed in our model, give us indirect hints of the physical phenomena associated with the specific heat maxima.
In general terms, the description of the nature of a phase transition, which usually manifests itself as sharp maximum in the specific heat data (see Fig.\,\ref{Fits-4levels}) in the low-temperature window,  as well as the physical mechanism responsible for the broad maximum in the high-temperature range require the combination of complementary experimental techniques.  Yet, it is worth mentioning that on the lower temperature flank, namely around the phase transition critical temperature $T_c$, the discrepancy between fits and the experimental data can be attributed to the absence of critical fluctuations, see e.g.\,Refs.\,\cite{PRL2010, PRL2008, 2015} and references cited therein, not negligible upon approaching $T_c$, in our \emph{Ansatz}. In the next section we discuss a degenerate Schottky model.

\section{Degenerate Schottky Model}
\begin{figure}
\centering
\includegraphics[angle=0,width=0.50\textwidth]{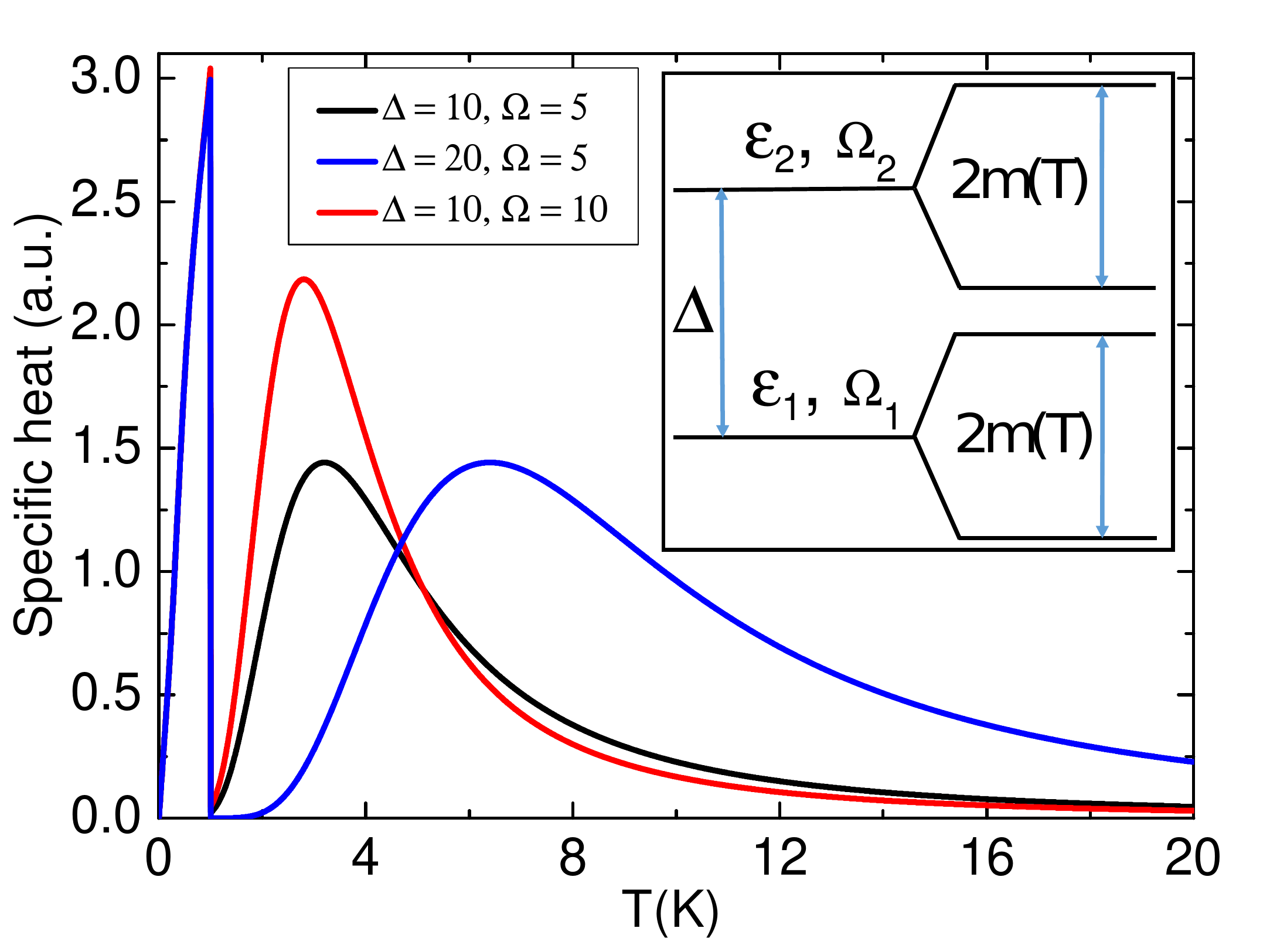}
\caption{\small Main panel: specific heat as a function of temperature for various energy gaps $\Delta$ and degeneracy ratio $\Omega$ = $\Omega_1$/$\Omega_2$, cf.\,indicated in the label. Inset: energy scheme of the degenerate Schottky model. See details in the main text.}
\label{degen-schott}
\end{figure}

Following the proposal of modelling and understanding of the double peak anomalies in specific heat data,
in this Section we focus on the description of a degenerate Schottky model. To this end, we consider a two level system $(\varepsilon _{1}=-\frac{1}{2}\Delta
,\varepsilon _{2}=\frac{1}{2}\Delta $, with  respective degeneracies $\Omega
_{1,2})$ plus a spin $\frac{1}{2}$ Zeeman splitting contribution $\pm \mu
_{B}H$, cf.\, scheme shown in the inset of Fig.\,\ref{degen-schott}. The partition function and equation for the energy (obtained using Eq.\,\ref{energia}) per particle is, disregarding an additive constant, given respectively by \cite{reif}:

\begin{eqnarray}
\label{Lagos-Z}
Z &=& \Omega_1 e^{-\beta \varepsilon_1}(e^{+\beta m_0 H} + e^{-\beta m_0 H}) +\nonumber \\& & \Omega_2 e^{-\beta \varepsilon_2}(e^{+\beta m_0 H} + e^{-\beta m_0 H}),
\end{eqnarray}

\begin{eqnarray}
\label{Lagos-1}
E &=&-mH-\frac{\Delta e^{-\beta \Delta }}{\left( 1+\Omega e%
^{-\beta \Delta }\right) },
\end{eqnarray}
where  $\Omega =\Omega _{1}/\Omega _{2}$, $H$ is an external magnetic field and  $%
\beta $ refers to the inverse temperature. The magnetization  $m$ and the specific heat per
particle are in turn given respectively by:

\begin{equation}
\label{Lagos-2}
m = \mu_Btanh(\beta\mu_BH),
\end{equation}

\begin{equation}
\label{Lagos-3}
C_V = \frac{\partial E}{\partial T} = -\frac{\partial}{\partial T}(mH) + \frac{\Omega(\beta\Delta)^2e^{-\beta\Delta}}{(1 + \Omega e^{-\beta\Delta})^2}.
\end{equation}

For an ordinary paramagnet, cf.\,discussion in Section\,I (see Eq.\,\ref{Schottky}), the magnetic contribution to the specific heat is Schottky type, namely it is described by the second term of Eq.\,\ref{Lagos-3} with  $\Omega = 1$ and an energy gap $\Delta = 2\mu _{B}H$, so that we have two decoupled Schottky like systems. We assume now that a ferromagnetic phase transition, which is a typical many body effect, takes place at a certain critical temperature $T_{%
\text{c}}$  within the Curie-Weiss molecular mean-field model \cite{reif}. We introduce then the macroscopic molecular field phenomenological parameter $\lambda$ and an effective magnetic field ($H_{eff}$), as follows:

\begin{eqnarray}
\label{Lagos-4}
H_{\text{eff}}=H_{\text{ext}}+\lambda m .
\end{eqnarray}%

At zero external magnetic field $m$ is calculated in a self-consistent fashion via
the following equation:

\begin{eqnarray}
\label{Lagos-5}
m=\mu _{B}\tanh (\beta \mu _{B}\lambda m),
\end{eqnarray}%
being thus now the magnetic gap $\Delta = 2\mu _{B}H=2\mu _{B}\lambda m$
 a function of the order parameter $m$. Scaling the magnetization to $\mu _{B},$ the energies (and temperatures) to $%
T_{\text{c}}$, with  $T_{\text{c}}\equiv \lambda \mu _{B}$, we have:

\begin{eqnarray}
\label{Lagos-6}
C_{V}=-\frac{\partial }{\partial T}\left( m^{2}\right) +\frac{\Omega (\beta
\Delta )^{2}e^{-\beta \Delta }}{\left(1+\Omega e^{-\beta
\Delta}\right)^{2}},
\end{eqnarray}%
where $m$ is the solution of $\ m=\tanh (\beta m)$. It is well known that $%
m\equiv 0$ for $T>1$ and $m(T)\neq 0$ otherwise. Also for small $m$ (for $T$
near and less than $1$) we have $m^{2}\approx 3(1-T)$ \ so $C_{V}$ exhibits
a jump $\delta C_{V}=3$ at $T=1$.
Fig.\,\ref{degen-schott} (main panel) depicts the specific heat as a function of temperature for various energy gaps $\Delta$ (in $T_{\text{c}}$ units) and degeneracy ratio $\Omega$, namely $\Delta$ = 10 and $\Omega$ = 5 (black solid line); $\Delta$ = 20 and $\Omega$ = 5 (blue solid line); $\Delta$ = 10 and $\Omega$ = 10 (red solid line) employing Eq.\,\ref{Lagos-5}. The low-$T$ peak, with a mean-field like shape, remains unaffected upon tuning $\Delta (>>1)$ and  $\Omega$. The Schottky maximum, however,  becomes broader and shifts to higher temperatures as $\Delta$ is increased, being the peak intensity also increased as $\Omega$ increases cf.\,expected for a two-level model, a feature that can be understood in terms of entropy arguments as discussed in a previous section. It is worth mentioning that between the ferromagnetic transition and Schottky maximum is exponentially small, i.e.\,the ferromagnetic phase transition produces a  jump in the specific heat data on top of a non-magnetic contribution.
For $\Delta$ values close to unity (not shown in Fig.\,\ref{degen-schott}) the high-$T$ maximum tail evolves into the low-$T$, being the sharp mean-field like jump $\Delta C = 3$ at $T = 1$ preserved. Thus, our analysis show that a simple Schottky model coupled to a ferromagnetic phase transition is also amenable to fit emerging double peak/maxima features in measured specific heat data.





\section{Conclusions}
To summarize, we have employed a multilevel model to describe specific heat data as a function of temperature with two maxima. Our
results suggest that: \emph{i}) a ratio of $\Delta_2$/$\Delta_1$
$\thickapprox$ 10 
between the energy levels and \emph{ii}) the energy levels degeneracy,
govern the two-peaks regime in the specific heat. 
Apart from the nature of the entities responsible for the emergence of the maxima in the specific heat data, this simple model describes nicely recent literature results for the CeAuGe compound. Furthermore, the specific heat of a degenerate Schottky model was calculated. Our analysis demonstrate that double peak/maxima features in specific heat data can also stem from a two-level model coupled to a ferromagnetic phase transition. We trust that our findings pave the way towards the description of experimental specific heat data for other systems in both soft and condensed matter Physics.


\section*{Acknowledgements}
MdS and RP acknowledge financial support from the S\~ao Paulo
Research Foundation -- Fapesp (Grants No. 2011/22050-4 and 2011/17253-3, respectively) and National Council of
Technological and Scientific Development -- CNPq (Grants No. 308977/2011-4 and 305472/2014-3).
\bibliography{References-SH1}
\end{document}